\documentclass[preprintnumbers,amsmath,amssymb,floatfix,11pt,prd,superscriptaddress,nofootinbib]{revtex4}
\usepackage{graphicx}
\usepackage{epsfig}
\usepackage{bm}
\usepackage{amsfonts}

\def\ben{\begin{equation}}
\def\een{\end{equation}}

 \def\bd{\begin{document}} \def\ed{\end{document}}
\def\ds{\documentstyle} \let\fr=\frac \let\bl=\bigl \let\br=\bigr
\let\Br=\Bigr \let\Bl=\Bigl
\let\bm=\bibitem
\let\na=\nabla
\let\pa=\partial \let\ov=\overline
\newcommand{\be}{\begin{equation}}
\newcommand{\ee}{\end{equation}}
\def\ba{\begin{array}}
\def\ea{\end{array}}
\def\ft#1#2{{\textstyle{\frac{\scriptstyle #1}{\scriptstyle #2} } }}
\def\fft#1#2{{\frac{#1}{#2}}}
\def\del{\partial}
\def\vp{\varphi}
\def\sst#1{{\scriptscriptstyle #1}}
\def\oneone{\rlap 1\mkern4mu{\rm l}}
\def\td{\tilde}
\def\wtd{\widetilde}
\def\ie{{\it i.e.\ }}
\def\dalemb#1#2{{\vbox{\hrule height .#2pt
        \hbox{\vrule width.#2pt height#1pt \kern#1pt
                \vrule width.#2pt}
        \hrule height.#2pt}}}
\def\square{\mathord{\dalemb{6.8}{7}\hbox{\hskip1pt}}}
\newcommand{\ho}[1]{$\, ^{#1}$}
\newcommand{\hoch}[1]{$\, ^{#1}$}
\newcommand{\bea}{\setlength\arraycolsep{2pt} \begin{eqnarray}}
\newcommand{\eea}{\end{eqnarray}}
\newcommand{\ra}{\rightarrow}
\newcommand{\lra}{\longrightarrow}
\newcommand{\Lra}{\Leftrightarrow}
\newcommand{\bp}{\tilde \beta^\prime}
\newcommand{\tr}{{\rm tr} }
\newcommand{\Tr}{{\rm Tr} }
\def\0{{\sst{(0)}}}
\def\1{{\sst{(1)}}}
\def\2{{\sst{(2)}}}
\def\3{{\sst{(3)}}}
\def\4{{\sst{(4)}}}
\def\5{{\sst{(5)}}}
\def\6{{\sst{(6)}}}
\def\7{{\sst{(7)}}}
\def\8{{\sst{(8)}}}
\def\m{{\sst{(m)}}}
\def\n{{\sst{(n)}}}
\def\cA{{{\cal A}}}
\def\cB{{{\cal B}}}
\def\cF{{{\cal F}}}
\def\cG{{{\cal G}}}
\def\cH{{{\cal H}}}
\def\tV{\widetilde V}
\def\tW{\widetilde W}
\def\tH{\widetilde H}
\def\tE{\widetilde E}
\def\tF{\widetilde F}
\def\tA{\widetilde A}
\def\im{{{\rm i}}}
\def\tY{{{\wtd Y}}}
\def\ep{{\epsilon}}
\def\vep{{\varepsilon}}
\def\bD{{{\bar D}}}
\def\R{{{\mathbb R}}}
\def\C{{{\mathbb C}}}
\def\H{{{\mathbb H}}}
\def\CP{{{\mathbb C}{\mathbb P}}}
\def\RP{{{\mathbb R}{\mathbb P}}}
\def\Z{{{\mathbb Z}}}
\def\bA{{{\mathbb A}}}
\def\bB{{{\mathbb B}}}
\def\bC{{{\mathbb C}}}
\def\bD{{{\mathbb D}}}
\def\bE{{{\mathbb E}}}
\def\bZ{{{\mathbb Z}}}
\def\Re{{{\frak{Re}}}}
\def\Im{{{\frak{Im}}}}
\def\cosec{{\,\hbox{cosec}\,}}
\def\Gm{{\Gamma_{\!\! -}}}
\def\Gp{{\Gamma_{\!\! +}}}
\def\stan{{standard }}
\def\nonstan{{supernumerary }}
\def\p{{\partial}}
\def\kdel#1{{\fft{\del}{\del#1}}}

\def\bog{{Bogomolny }}
\def\om{{\omega}}
\newcommand{\un}{\underline}
\def\R{\hbox{{\rm I}\kern-0.2em{\rm R}\kern0.2em}}%mathematical R for reals
\def\D{\hbox{{\rm I}\kern-0.2em{\rm D}\kern0.2em}}
\def\a{\alpha} \def\o{\omega} \def\w{\wedge}
\def\b{\beta}         \def\rf{\rfloor}
\def\e{{\rm e}}
\def\ld{\lambda} \def\Ld{\Omega}
\def\d{{\rm d}}
\def\dsy{\displaystyle}
\def\de{\delta}
\def\ep{\epsilon}
\def\g{\gamma}
\def\be{\begin{equation}}
\def\ee{\end{equation}}
\def\X{{\cal X}} \def\U{{\cal U}}
\def\p{\partial}
\def\({\left(}
\def\){\right)}
\def\[{\left[}
\def\]{\right]}
\def\bc{\begin{center}}
\def\ec{\end{center}}
\def\EL{Euler-Lagrange}
\def\ph{\phantom}
\def\wh{\widehat}
\def\n{\noindent}
\newcommand{\nnr}{\nonumber \\}
\newcommand{\pd}{\partial}
\newcommand{\ud}{\textrm{d}}
\newcommand{\dTH}{T^{\prime \, 0}_\textrm{H}}
\newcommand{\dOi}{\Omega^{\prime \, 0}_i}
\newcommand{\bx}{{\bf x}}
\begin{document}

\title{Noether Symmetry Approach in $f(R)-$Tachyon Model}
\author{\textbf{Mubasher Jamil}}
\email{mjamil@camp.nust.edu.pk} \affiliation{Center for Advanced
Mathematics and Physics (CAMP), National University of Sciences and
Technology (NUST), H-12, Islamabad, Pakistan}

\author{\textbf{Fazal M. Mahomed}}
\email{Fazal.Mahomed@wits.ac.za} \affiliation{Centre for
Differential Equations, Continuum Mechanics and Applications, School
of Computational and Applied Mathematics, University of the
Witwatersrand, Wits 2050, South Africa}

\author{\textbf{D. Momeni}}
\email{d.momeni@yahoo.com} \affiliation{Department of Physics,
Faculty of Sciences, Tarbiat Moa'llem University, Tehran, Iran}

\begin{abstract}
\vspace*{1.5cm} \centerline{\bf Abstract} \vspace*{1cm}

In this Letter by utilizing the Noether symmetry approach in
cosmology, we attempt to find the tachyon potential via the
application of this kind of symmetry to a flat
Friedmann-Robertson-Walker (FRW) metric. We reduce the system of
equations to simpler ones and obtain the general class of the
tachyon's potential function and $f(R)$ functions.  We have found
that the Noether symmetric model results in a power law $f(R)$ and
an inverse fourth power potential for the tachyonic field. Further
we investigate numerically the cosmological evolution of our model
and show explicitly the behavior of the equation of state crossing
the cosmological constant boundary.
\end{abstract}

\maketitle

\newpage

\section{Introduction}

Observations of type Ia supernovae (SNIa) indicate that currently
the observable Universe is undergoing an accelerating expansion
\cite{riess}. This cosmic acceleration has also been confirmed by
numerous observations of large scale structure (LSS) \cite{3} and
measurements of the cosmic microwave background (CMB) anisotropy
\cite{4}. The cause of this cosmic acceleration is generally labeled
as ``dark energy'', a mysterious exotic energy which generates a
large negative pressure, whose energy density  dominates the
Universe (for a review see e.g. \cite{sami}). The astrophysical
nature of dark energy is that it does not cluster at any scale
unlike normal baryonic matter which forms structures. The combined
analysis of cosmological observations suggests that the Universe is
spatially flat and consists of about 70\% dark energy, 30\% dust
matter (cold dark matter plus baryons) and negligible radiation. The
nature of dark energy as well as its cosmological origin remain
mysterious at present.

One of the approaches to the construction of a dark energy model is
to modify the geometrical part of the Einstein equations. The
general paradigm consists in adding into the effective action,
physically motivated higher-order curvature invariants and
non-minimally coupled scalar fields. The representative models based
on this strategy are termed `modified gravity' and include $f(R)$
gravity \cite{fr}, Horava-Lifshitz gravity \cite{hl,hl2,hl3},
scalar-tensor gravity \cite{st,st1} and the braneworld model
\cite{brane,brane1}. Modified Gravity has been successful to explain
the rotation curves of galaxies, the motion of galaxy clusters and
the Bullet Cluster \cite{mofat}.

Besides compatibility with the observational data, the minimal
criteria that a modified gravity theory must satisfy in order to be
viable are \cite{setare}: (1) reproducing the desired dynamics of
the Universe including an inflationary era, followed by a radiation
era and a matter era and finally, by the present acceleration epoch;
(2) the theory must have Newtonian and post-Newtonian limits
compatible with the available solar system observational data; (3)
the theory must not have deviations from general relativity at the
level of accuracy following from present laboratory and solar system
tests of gravity; (4) the theory must possess a future stable (or at
least meta-stable) de Sitter asymptote, which is necessary for a
description of the present dark energy; (5) the theory must be
stable at the classical and quantum level.

The $f(R)$ theory of gravity is a meticulous class of modified
theories of gravity. This theory can be obtained by replacing the
Ricci scalar $R$ with an arbitrary function $f(R)$ in the
Einstein-Hilbert Lagrangian. The dynamical equations of motion can
be obtained by varying the Lagrangian with respect to the metric
(metric formalism) or viewing the metric and connections as
independent variables and varying the action with respect to both
independently (Platini formalism) \cite{voll,Meng}. Nojiri and
Odintsov have shown that inflation and current cosmic acceleration
may take place by adding positive and negative powers of curvature
into the Einstein-Hilbert Lagrangian \cite{NO}. Carroll et al. have
proposed that by adding an inverse term of $R$ to the
Einstein-Hilbert Lagrangian would lead to cosmic speed-up which will
instigate purely gravitational effects \cite{carr}. It should be
mentioned that the main deficiency of such theories is that they are
solemnly constrained by solar system tests \cite{Olm,chi}. Amendola
et al. \cite{Amen} and Starobinsky \cite{Star} have proposed
different forms of $f(R)$ that can satisfy both cosmological and
local gravity constraints.

In the past, the use of tachyon in certain string theories has been
explored which has resulted in a better understanding of the D-brane
decaying process \cite{sen,new}. This led to study the role of
tachyon in cosmology as well. A rolling tachyon field $\phi$ has an
equation of state whose parameter smoothly interpolates between $-1$
and $0$ \cite{a}. Thus, a tachyon can be realized as a suitable
candidate for the inflation at high energy \cite{b} as well as a
source of dark energy depending on the form of the tachyon potential
\cite{c}. Therefore it becomes meaningful to reconstruct a tachyon
potential $V(\phi)$ in the framework of $f(R)$ gravity. It was
demonstrated that dark energy driven by tachyon, decays to cold dark
matter in the late accelerated Universe and this phenomenon yields a
solution to the cosmic coincidence problem \cite{d}.

The plan of this letter is as follows: In Section II, we present the
formal framework of the $f(R)$-Born-Infeld effective action of
tachyon. In section III, we construct the governing differential
equations from the Noether condition and solve them in an
accompanying subsection. In section IV, we study the dynamics of the
present model. Finally we conclude this work.

\section{ Formal framework of the $f(R)$-Born-Infeld effective action of tachyon}

We consider a spatially flat FRW cosmology with a tachyon part is
taken as the usual Born-Infeld action. A generalization of the
Einstein-Hilbert action with a modified tachyon action for matter
sector has been discussed previously \cite{cai2007}. The action in
$(n+1)$  dimensions is
%%%%
\be S=\int
d^{n+1}x\sqrt{-g}[f(R)-V(\phi)\sqrt{1-\alpha'\nabla_{\mu}\phi\nabla^{\mu}\phi}].
\label{1} \ee
 %%%
We take $c=1,16\pi G=1$, $sig(g)=1-n$ and the coordinates are
$x^{\mu}=(t,x^{i}),i=2...n+1$. We define
$\alpha'=\frac{\alpha}{M^4}$ as the coupling constant and $M$ an
energy scale to make the kinetic part of the action dimensionless.
For $n=3$, the action (\ref{1}) represents the 4-D effective action
of tachyon field and gives the dynamics to the lowest order in
$\nabla_{\mu}\phi\nabla^{\mu}\phi$. The function $f(R)$ is an
arbitrary function of the Ricci scalar $R$. The energy-momentum (EM)
tensor for the tachyon field is \cite{EM}
%%%%
\be T_{\text{T}}^{\mu\nu}=g^{\mu\nu}V(\phi)h
+\frac{\alpha}{M^4}\frac{V(\phi)}{h}
\nabla^{\mu}\phi\nabla^{\mu}\phi. \label{2}
 \ee
 %%%
Here we take $h=\sqrt{1-\alpha'\nabla_{\mu}\phi\nabla^{\mu}\phi}$.
By varying the action (\ref{1}) with respect to the metric
$g_{\mu\nu}$ and the scalar field $\phi$, we obtain the
corresponding equations of motion (EOM):\footnote{In this paper we
adopt $\dot{a}=\frac{d a}{d t},f'=\frac{\partial f}{\partial R}$}:
%%%
\be
 \frac{1}{2}g_{\mu\nu}f(R)-f'(R)R_{\mu\nu}+\nabla_{\mu}\nabla_{\nu}f'(R)
 -g_{\mu\nu}\square
 f'(R)=\frac{1}{2}\Big(g_{\mu\nu}V(\phi)h+
 \frac{\alpha}{M^4}\frac{V(\phi)}{h}
\nabla^{\mu}\phi\nabla^{\mu}\phi\Big),\label{3}
 \ee
  %%%
  \be
\nabla_{\mu}\Big(\frac{V(\phi)\nabla^{\mu}\phi}{h}\Big)+\frac{h}{\alpha'}\frac{dV(\phi)}{d\phi}=0.
\label{4} \ee
%%%
Our main goal is the construction of a potential function $V(\phi)$
and the exact form of the gravity sector $f(R)$ using the Noether
symmetry by following the procedure of \cite{Noether}. If the
tachyon sector is removed, the resulting action is nothing but the
$f(R)$ action whose symmetry analysis (without the gauge term) has
been discussed in \cite{babak}.

\section{Noether symmetry approach in $f(R)$-tachyon model in four dimensions}

We consider the action (\ref{1}) representing the dynamical system
in which the scale factor  $a(t)$, curvature scalar $R$ and the
tachyon field $\phi$ play the role of independent dynamical
variables. We can write (\ref{1}) in a background of flat FRW metric
$g_{\mu\nu}=diag(1,-a^2(t)\eta_{ij})$, ($i,j=2,3,4$) as
%%%
  \be
S=\int dt\Big[a^3\Big(f(R)-V(\phi)\sqrt{1-\alpha'\dot{\phi}^2}\Big)-
\lambda\Big(R-6\Big(\Big(\frac{\dot{a}}{a}\Big)^2+\frac{\ddot{a}}{a}\Big)\Big)\Big].
\label{5} \ee
%%%
We can obtain the Lagrange-multiplier $\lambda$ by varying the
action (\ref{5}) with respect to $R$. This procedure leads to
$\lambda=a^3 f'(R)$. For a purely vacuum $f(R)$-tachyon cosmology,
we obtain the following Lagrangian
%%%
  \be
L(a,\dot{a},R,\dot{R},\phi,\dot{\phi})=6\dot{a}^2af'+6\dot{a}\dot{R}a^2f''+a^3(f'R-f)
-a^3V(\phi)\sqrt{1-\alpha'\dot{\phi}^2} \label{6}. \ee

\subsection{Exact solutions}

Noether symmetries are the symmetries associated with Lagrangians
which may help in discovering new features of the gravitational
theories. For instance, the application of Noether symmetries in
higher-order theory of gravity turns out to be a powerful tool to
find the solution of the field equations \cite{sanyal}. The Noether
symmetry approach when applied to scalar-tensor cosmology yields an
extra correction term $R^{-1}$ and fixes the form of the coupling
parameter and the field potential \cite{capo}. Noether symmetries
when applied to a generic $f(R)$ cosmological model yields exact
forms of the $f(R)$ functions and also generates an effective state
parameter that produces cosmic acceleration
\cite{babak,babak1,babak2,capo1,capo11}. A similar approach when
applied to Platini $f(R)$ gravity yields a power-law form $f(R)\sim
R^n$ \cite{roshan}. Recently a model-independent criterion has been
proposed based on first integrals of motion, due to Noether
symmetries of the equations of motion, in order to classify the dark
energy models in the context of scalar field (quintessence or
phantom) FRW cosmologies \cite{new1}. Although in the literature
Noether symmetries have been studied in the context of $f(R)$ theory
of gravity \cite{babak,babak1,babak2,capo1,capo11}, all these
authors have used the definition of Noether symmetries without a
{\it gauge} term. Taking into account the gauge term gives a more
general definition \cite{Ibr,Ibr1} of the Noether symmetries. Thus
one may expect some extra symmetry generators from this definition
and hence one may obtain some extra (new) forms of $f(R)$. Here we
apply the Noether condition with the gauge term to look at some
interesting forms of $f(R)$.

A vector field
\begin{equation}\label{13}
X={\tau}(t,a,R,\phi){\p\over\p t}+\a(t,a,R,\phi){\p\over\p
a}+\b(t,a,R,\phi) {\p\over\p R}+\g(t,a,R,\phi){\p\over\p \phi},
\end{equation}
is a Noether symmetry corresponding to a Lagrangian
$L(t,a,R,\phi,\dot a,\dot R, \dot\phi)$ if
\begin{equation}\label{14}
X^{[1]}L+LD_t(\tau)=D_tB,
\end{equation}
holds, where $X^{[1]}$ is the first prolongation of the generator
$X$, $B(t,a,R,\phi)$ is a gauge function and $D_t$ is the total
derivative operator
\begin{equation}\label{14a}
D_t\equiv\frac{\partial }{\partial t}+\dot a\frac{\partial
}{\partial a}+ \dot R\frac{\partial}{\partial R}+\dot
\phi\frac{\partial }{\partial \phi}.
\end{equation}
The prolonged vector field is given by
\begin{equation}\label{15}
X^{[1]}=X+\a_t{\p\over\p \dot a}+\b_t{\p\over\p \dot
R}+\g_t{\p\over\p \dot\phi},
\end{equation}
in which
\begin{equation}\label{16}
\a_t=D_t\a-\dot aD_t\tau, \ \ \b_t=D_t\b-\dot R D_t\tau,\ \
\g_t=D_t\g-\dot \phi D_t\tau.
\end{equation}
The Noether condition (8) results in the over-determined system of
equations
\begin{eqnarray}
\g(\phi)'-\dot\tau(t)&=&0,\label{17}\\
3\a V+\g a V'+\dot\tau a V&=&0,\label{18}\\
\a_R=\a_\phi&=&0,\label{19}\\
\b_\phi&=&0,\label{20}\\
\a f'+\b a f''+2 a f'\a_a- a f'\dot\tau+  a^2f''\b_a&=&0,\label{21}\\
2a\a f''+a^2\b f'''+a^2 \a_a f''+a^2f'' \b_R&=&0,\label{22}\\
12\a_taf'+6a^2 f''\b_t&=&B_a,\label{23}\\
6a^2f''\a_t&=&B_R,\label{24}\\
B_\phi&=&0,\label{25}\\
(3a^2\a+\dot\tau a^3)(f'R-f)+\b a^3 f''R&=&B_t,\label{26}
\end{eqnarray}
provided $f''\ne0$. Eq. (\ref{17}) implies
\begin{eqnarray}
\g=c_1\phi+c_2,\nonumber\\
\tau=c_1t+c_3, \label{27}
\end{eqnarray}
where $c_i$s are constants. Then Eqs. (\ref{18}) and (\ref{19})
 give \be \a=c_4a, \label{28} \ee where $c_4$ is a further
arbitrary constant. Thus $V(\phi)$ satisfies the ordinary
differential equation \be (3c_4+c_1)V+(c_1\phi+c_2)V'=0. \label{29}
\ee Its solution is \be V=V_0(\phi+\phi_0)^{-4}, \nonumber \ee where
$V_0$ and $\phi_0$ are constants. Eqs. (\ref{20}),
(\ref{23})-(\ref{26}) and (\ref{28}) further reveal that $\b$
satisfies \be \b f''R+(3c_4+c_1)(f'R-f)=c_5a^{-3},\label{30} \ee
where $c_5$ is a constant. Then (\ref{21}) gives rise to $f(R)$
being of the form \be f(R)=rR^\nu,\label{31} \ee where $r$ is a
constant and $\nu=(3c_4+c_1)/2c_1$ provided $c_1\ne0$. Note that
$c_1=0$ results in $f$ being constant so it is excluded from further
consideration. Also $c_5$ in (\ref{30}) turns out to be zero as a
consequence of (\ref{21}).

The insertion of (\ref{31}) into (\ref{26}) yields \be \b=-2c_1R.
\ee

Eq. (\ref{22}) now provides the further constraint \be c_4=c_1. \ee
As we saw earlier $\nu=(3c_4+c_1)/2c_1$, thus using (28) we deduce
$\nu=2$ and therefore the quadratic power law \be f(R)=rR^2. \ee

For $f(R)=rR^2$ and $V=V_0(\phi+\phi_0)^{-4}$, there are two Noether
symmetries given by
\begin{eqnarray}
X_1&=&{\p\over\p t},\nonumber\\
X_2&=&t{\p\over\p t}+a{\p\over\p a}+(\phi+\phi_0){\p\over\p
\phi}-2R{\p\over\p R}.
\end{eqnarray}
Here the gauge function is zero. The first symmetry $X_1$
(invariance under time translation) gives the energy conservation of
the dynamical system in the form of (31) below, while the second
symmetry $X_2$ (scaling symmetry) and a corresponding conserved
quantity of the form (32) below. The two first integrals (conserved
quantities) which are
\begin{eqnarray}
I_1&=&\tau L-\dot a{\p L\over \p \dot a}-\dot R{\p L\over\p \dot R}
-\dot \phi{\p L\over\p \dot \phi},\nonumber\\
&=& -6a\dot{a}^2f'-6a^2\dot a \dot
Rf''+a^3(f'R-f)-a^3V\sqrt{1-\alpha'\dot \phi^2}\nonumber\\&&
-\alpha'a^3 V\dot\phi^2(1-\alpha'\dot\phi^2)^{-1/2}.\\
I_2&=&tL+(a-t\dot a){\p L\over\p \dot a}+(-2R-t\dot R){\p L\over\p \dot R}
+(\phi+\phi_0-t\dot\phi){\p L\over\p \dot \phi},\nonumber\\
&=&-12a{\dot a}^2trR+a^3trR^2+12a^3r\dot R-12a^2t\dot a\dot R-ta^3
\sqrt{1-\a'{\dot \phi}^2}V_0(\phi+\phi_0)^{-4}\nonumber\\
&&-\a'a^3t{\dot \phi}^2(1-\a'{\dot
\phi}^2)^{-1/2}V_0(\phi+\phi_0)^{-4}+\a'a^3\dot\phi
V_0(\phi+\phi_0)^{-3} (1-\a'{\dot \phi}^2)^{-1/2}.
\end{eqnarray}

\section{Cosmic evolution}

According to the observations of type Ia supernovae Gold dataset
\cite{c,gold}, there exists the possibility that the effective
equation of state (EOS) parameter, which is the ratio of the
effective pressure of the Universe to the effective energy density,
evolves from values greater than $-1$ to less than $-1$ (see
\cite{noji} for extensive set of references on the studies of
phantom crossing in different frameworks), namely, it crosses the
cosmological constant boundary (the phantom divide) currently or in
near future. In this section, we derive the effective equation of
state that admits the phantom crossing with suitable adjustment of
parameters.

The field equation (3) can be rewritten in the form of Einstein
equations with an effective stress-energy tensor. Specifically, as
\begin{eqnarray}
G_{\mu\nu}&=&\kappa(T^{\text{T}}_{\mu\nu}+T^{\text{eff}}_{\mu\nu})\\\nonumber
&=&\kappa\Big(g^{\mu\nu}V(\phi)h +\alpha'\frac{V(\phi)}{h}
\nabla^{\mu}\phi\nabla^{\nu}\phi+\frac{f(R)-Rf'(R)}{2}g_{\mu\nu}+\nabla_{\mu}\nabla_{\nu}f'(R)-g_{\mu\nu}\Box
f'(R)\Big).
\end{eqnarray}
Here $\kappa=\frac{1}{2}$. Since $T^{\text{eff}}_{\mu\nu}$ is only a
formal energy-momentum tensor, it is not expected to satisfy any of
the energy conditions deemed reasonable for physical matter, in
particular the effective energy density cannot be expected to be
positive-definite. An effective gravitational coupling
$G_{\text{eff}} = \frac{G}{f'(R)}$ can be defined in a way analogous
to scalar-tensor gravity. It is apparent that $f'(R)$ must be
positive for the graviton to carry positive kinetic energy.
Motivated by recent cosmological observations, we adopt the
spatially flat FRW metric
\begin{eqnarray}
ds^2=-dt^2+a^2(t)(dx^2+dy^2+dz^2).
\end{eqnarray}
Then the field equations for the $f(R)$-tachyon cosmology become
\begin{eqnarray}
H^2&=&\frac{1}{6f'(R)}\Big(\rho_{\text{T}}+\frac{f(R)-Rf'(R)}{2}-3H\dot{R}f''(R)\Big),\\
2\dot{H}+3H^2&=&-\frac{1}{2f'(R)}\Big(P_\text{T}+f'''(R)\dot{R}^2+2H\dot{R}f''(R)+\ddot{R}f''(R)
+\frac{f(R)-Rf'(R)}{2}\Big),
\end{eqnarray}
For $f(R)=rR^2$ and $V=V_0(\phi+\phi_0)^{-4}$, we have
\begin{eqnarray}
H^2&=&\frac{1}{12rR}\Big(V_0r(\phi+\phi_0)^{-4}(1+\alpha'{\dot
\phi}^2)^{1/2}-\alpha'
 \frac{V_0(\phi+\phi_0)^{-4}}{(1+\alpha'{\dot \phi}^2)^{1/2}}\dot{\phi}^2-\frac{rR^2}{2}-
 6rH\dot{R}\Big),\\
2\dot{H}+3H^2&=&-\frac{1}{4rR}\Big(-V_0r(\phi+\phi_0)^{-4}(1+\alpha'{\dot
\phi}^2)^{1/2}+4rH\dot{R}+2r\ddot{R}-\frac{rR^2}{2}\Big).
\end{eqnarray}
Thus
\begin{eqnarray}
\rho_{\text{tot}}&=&\frac{1}{2rR}\Big(V_0r(\phi+\phi_0)^{-4}(1+\alpha'{\dot
\phi}^2)^{1/2}-\alpha'\dot{\phi}^2
 \frac{V_0(\phi+\phi_0)^{-4}}{(1+\alpha'{\dot
 \phi}^2)^{1/2}}-\frac{rR^2}{2}-6rH\dot{R}\Big),\\
 P_{\text{tot}}&=&\frac{1}{2rR}\Big(-V_0r(\phi+\phi_0)^{-4}(1+\alpha'{\dot
\phi}^2)^{1/2}+4rH\dot{R}+2r\ddot{R}-\frac{rR^2}{2}\Big).
\end{eqnarray}
Hence, the effective equation of state parameter for the
$f(R)$-tachyon cosmology is
\begin{eqnarray}
w_{\text{eff}}=\frac{P_{\text{tot}}}{\rho_{\text{tot}}}=\frac{-V_0r(\phi+\phi_0)^{-4}(1+\alpha'{\dot
\phi}^2)^{1/2}+4rH\dot{R}+2r\ddot{R}-\frac{rR^2}{2}}{V_0r(\phi+\phi_0)^{-4}(1+\alpha'{\dot
\phi}^2)^{1/2}-\alpha'\dot{\phi}^2
 \frac{V_0(\phi+\phi_0)^{-4}}{(1+\alpha'{\dot
 \phi}^2)^{1/2}}-\frac{rR^2}{2}-6rH\dot{R}}.
\end{eqnarray}
For simplicity, we take $V_0=r=\alpha'=1,\phi_0=0$; therefore the EOS is
now
\begin{eqnarray}
w_{\text{eff}}=\frac{-\phi^{-4}(1+{\dot
\phi}^2)^{1/2}+4H\dot{R}+2\ddot{R}-\frac{R^2}{2}}{\phi^{-4}(1+{\dot
\phi}^2)^{1/2}-
 \frac{\phi^{-4}\dot{\phi}^2}{(1+{\dot
 \phi}^2)^{1/2}}-\frac{R^2}{2}-6H\dot{R}}.
\end{eqnarray}
\begin{figure}
\centering
 \includegraphics[scale=0.5]{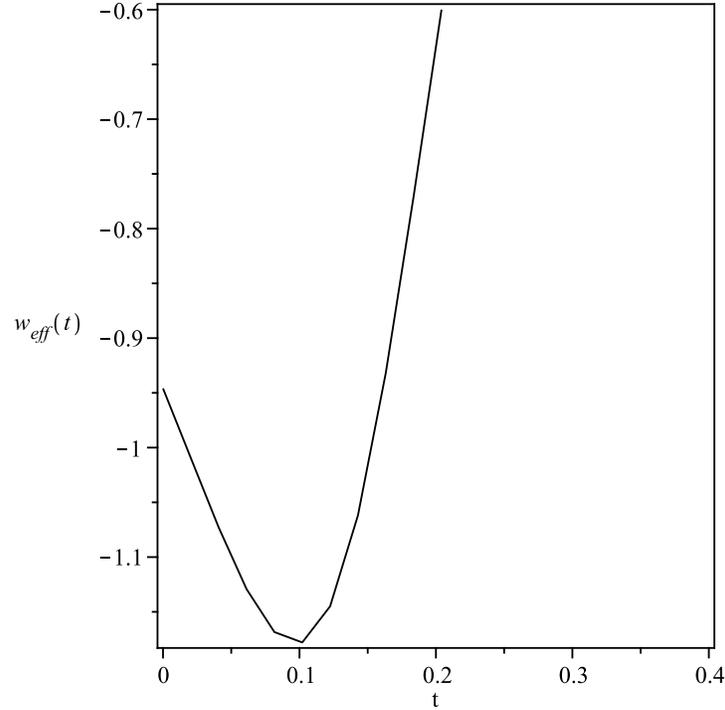} % scale goes from 0 to 1.
  \caption{ The  general behavior of the $w_{eff}$ for a set
  of initial conditions. It shows the phantom
crossing line $w_{eff}=-1$.
  }
 \label{fig5.eps}
\end{figure}
For better understanding of this type of phase transition we must
analyze (42). For metric (34), the equations of motion for scalar
field $\phi(t)$ and the scale factor $a(t)$ are
\begin{equation}
\frac{1}{a^3}\frac{d (a^3\dot{\phi})}{dt}-\frac{d\log
h}{dt}\dot{\phi}+\frac{4}{\alpha' \phi}=0,
\end{equation}
\begin{equation}
\Big(\frac{\dot{a}(t)}{a(t)}\Big)^2=\frac{1}{12R}\Big(\phi^{-4}(1+\alpha'{\dot
\phi}^2)^{1/2}-
 \frac{\alpha'\dot{\phi}^2\phi^{-4}}{(1+\alpha'{\dot \phi}^2)^{1/2}}-\frac{R^2}{2}-
 6H\dot{R}\Big).
\end{equation}
Here $R$ is the Ricci scalar of metric (34). We solved Eqs. (43) and
(44) for $a(t)$ and $\phi(t)$ numerically for a suitable set of the
initial conditions imposed on these functions. Fig.(1) shows the
general behavior of the $w_\text{eff}$. It shows the phantom
crossing line $w_\text{eff}=-1$. It can crosses the dark energy line
$w=-1$ several times with respect to the values of $H,\phi,R$.
Further, since $-1 < w_{\text{eff}} < 0$, one type of dark energy,
namely quintessence, can be addressed in the model described above.
Also another type of dark energy known as phantom, with
$w_{\text{eff}} < -1$, can be accounted.

\section{Conclusion}

In this work, we have studied the $f(R)$-tachyon cosmology by the
Noether symmetry approach. This approach is based on the search for
Noether symmetries which allow one to find the form of the function
$f(R)$ and the tachyon's potential $V=V(\phi)$. We have shown that
the Noether symmetric model results in a power law expansion
$f(R)=rR^2$ for the action (up to a constant multiplicative factor)
and an inverse fourth power $V=V_0(\phi+\phi_0)^{-4}$ for tachyon's
potential. This form may be of interest to tachyonic cosmology and
it's extensions. Moreover, the gauge function turns out to be zero.
The case $V$=constant was not considered for which the Noether
symmetry is translation in phi or something equivalent. Also, by
analyzing the equation of state, we addressed the so-called crossing
the phantom divide line ($w = -1$) of dark energy.

\subsection{Acknowledgments}
The authors are grateful to the anonymous referee for useful
suggestions. They also thank M. Umar Farooq and M. Safdar for
insightful discussions related to this work.

\end{document}